\documentclass[aps,prl,twocolumn,superscriptaddress,showpacs]{revtex4}

\usepackage{bm}        
\usepackage{color}        

\usepackage{epsfig,graphicx,amsmath}

\def\percc{cm\ensuremath{^{-3}}}

\begin{document}


\title{Cold N+NH Collisions in a Magnetic Trap}


\author{Matthew T. Hummon}
\email{matt@cua.harvard.edu}
\affiliation{Department of Physics, Harvard University, Cambridge, MA 02138}
\affiliation{Harvard-MIT Center for Ultracold Atoms, Cambridge, MA 02138}

\author{Timur V. Tscherbul}

\affiliation{Harvard-MIT Center for Ultracold Atoms, Cambridge, MA 02138}
\affiliation{ITAMP, Harvard-Smithsonian Center for Astrophysics, Cambridge, MA 02138}

\author{Jacek K{\l}os}
\affiliation{Department of Chemistry and Biochemistry, University of Maryland, College Park, Maryland 20742}

\author{Hsin-I Lu}
\affiliation{School of Engineering and Applied Sciences, Harvard University, Cambridge, MA 02138}
\affiliation{Harvard-MIT Center for Ultracold Atoms, Cambridge, MA 02138}

\author{Edem Tsikata}
\affiliation{Department of Physics, Harvard University, Cambridge, MA 02138}
\affiliation{Harvard-MIT Center for Ultracold Atoms, Cambridge, MA 02138}

\author{Wesley C. Campbell}
\affiliation{Department of Physics, Harvard University, Cambridge, MA 02138}
\affiliation{Harvard-MIT Center for Ultracold Atoms, Cambridge, MA 02138}

\author{Alexander Dalgarno}
\affiliation{Harvard-MIT Center for Ultracold Atoms, Cambridge, MA 02138}
\affiliation{ITAMP, Harvard-Smithsonian Center for Astrophysics, Cambridge, MA 02138}

\author{John M. Doyle}
\affiliation{Department of Physics, Harvard University, Cambridge, MA 02138}
\affiliation{Harvard-MIT Center for Ultracold Atoms, Cambridge, MA 02138}


\date{\today}

\begin{abstract} We present an experimental and theoretical study of atom-molecule collisions in a mixture of cold, trapped atomic nitrogen and NH molecules at a temperature of $\sim 600$~mK.  We measure a small N+NH trap loss rate coefficient of $k^{(\mathrm{N+NH})}_\mathrm{loss} = 8(4) \times 10^{-13}$~cm$^{3}$s$^{-1}$. Accurate quantum scattering calculations based on {\it ab initio} interaction potentials are in agreement with experiment and indicate the magnetic dipole interaction to be the dominant loss mechanism. Our theory further indicates the ratio of N+NH elastic to inelastic collisions remains large ($>100$) into the mK regime.
\end{abstract}


\maketitle

The study of low temperature molecular collisions and interactions is a rapidly expanding area of research at the interface of chemistry and physics \cite{Carr:2009oz,Krems:2008pccp}.  Experimental and theoretical techniques have been developed for studying a wide variety of phenomena, ranging from quantum threshold scattering \cite {Gilijamse_OH_Xe,Sawyer:2008qq}, inelastic atom - molecule collisions \cite{campbell2009mechanism, Zahzam:2006la,Staanum:2006it}, external field control of dipolar interactions in cold and ultracold molecules \cite{Sawyer:2010mb, Ni:2010xy}, to chemistry at cold \cite{PhysRevLett.100.043203, Cresu:06ag,PhysRevLett.105.033001} and ultracold temperatures \cite{Krems:2008pccp,Ospelkaus:2010sci}. The immense diversity of molecular structure and interactions is the cornerstone of many applications of cold molecules in quantum information science \cite{DeMille:2002xr}, condensed-matter physics \cite{barnett:190401}, precision measurement \cite{Hudson:02ed,Vutha:10jp}, cold controlled chemistry \cite{Softley:09mp}, and astrophysics \cite{Akyilmaz_NO_depletion}.  Achievement of these applications relies on improved methods for cooling and further understanding of low temperature molecular collisions.

Recently, several experimental techniques were developed to study molecular collisions at cold and ultracold temperatures, beyond the reach of traditional molecular beam experiments \cite{Scoles}.  Ultracold ground state molecules with temperatures below 1 $\mu$K can be created via coherent state transfer from magneto-asssociated ultracold atomic gases, producing KRb \cite{Ni:2008eq} and Cs$_2$ \cite{Danzl:2010cs}.  Alternatively, direct cooling techniques such as buffer-gas cooling \cite{Hummon:2008rm} and Stark deceleration \cite{Gilijamse_OH_Xe} provide more general methods for cooling a larger class of molecules, enabling the production of cold molecules starting from room-temperature sources as an input, such as supersonic beams \cite{Gilijamse_OH_Xe,Krems:2008pccp,Carr:2009oz}.  Since the maximum molecular trap densities for these direct cooling techniques are, so far, $10^8-10^9$~cm$^{-3}$,  the collision experiments with these molecular samples typically use a dense rare gas collision partner \cite{campbell2009mechanism, Gilijamse_OH_Xe} or perform collisions at energies $> 10$~cm$^{-1}$ \cite{Gilijamse_OH_Xe,Sawyer:2008qq,Cresu:06ag}.  Further cooling and compression of molecular samples produced via direct cooling techniques is key to study cold molecule-molecule interactions and their applications. One possible approach is to sympathetically cool a molecule via collisions with an atom \cite{Carr:2009oz,Lara_Rb_OH,zuchowski:022701, Tacconi_NH_Rb,Wallis:2009ul,Barker:09nj}. Thus, studies of cold atom-molecule collisions not only uncover the underlying collision physics, but also may lead to important new methods for the production of a variety of ultracold molecules.



A number of theoretical studies have focused on the optimal selection of atomic collision partners for sympathetic cooling of molecules \cite{Lara_Rb_OH,zuchowski:022701, Tacconi_NH_Rb,Wallis:2009ul,Barker:09nj}. Almost all of these studies focused on the alkali-metal atoms, which can be easily prepared at milli-Kelvin temperatures via laser cooling.  
However, accurate quantum scattering calculations have shown that collision-induced inelastic relaxation of molecules such as OH \cite{Lara_Rb_OH}, ND$_3$ \cite{zuchowski:022701} and NH \cite{Tacconi_NH_Rb} with the alkali-metal atoms occurs rapidly, causing molecular trap loss and severely limiting the efficiency of sympathetic cooling. A recent theoretical study suggested that sympathetic cooling of NH molecules by collisions with laser-cooled alkaline earth atoms such as Mg might be possible at low magnetic field strengths \cite{Wallis:2009ul}. A related study indicated the possibility of sympathetic cooling of large molecules by collisions with rare-gas atoms in an optical dipole trap \cite{Barker:09nj}. We have recently suggested that cold atomic nitrogen (N) can be used for sympathetic cooling of open-shell molecules, and demonstrated co-trapping of N atoms with NH molecules \cite{Hummon:2008rm}, but the presence of $^3$He buffer gas at 0.5 K precluded measurement of N+NH collisions in that experiment. 

In this Letter, we report the observation of cold (570 mK) collisions between magnetically trapped, ground state open-shell atoms and ground-state polar molecules, N and NH, both species spin polarized. We measure a small N+NH trap loss rate coefficient of $k^{(\mathrm{N+NH})}_\mathrm{loss} = 8(4) \times 10^{-13}$~cm$^{3}$s$^{-1}$. To interpret experimental observations, we carry out accurate {\it ab initio} and quantum scattering calculations of N+NH collisions in a magnetic field. Our theoretical results agree with the observations and predict high elastic-to-inelastic ratios ($\gamma$) over a wide range of temperatures (from 1K down to about 1 mK), suggesting the feasibility of sympathetic cooling of NH molecules by collisions with N atoms in a magnetic trap. Recent unpublished theory work by another group has made similar theoretical claims \cite{Zuchowski:2010bh}. 

The apparatus used to co-trap atomic nitrogen and NH is similar to that described in our previous work \cite{Hummon:2008rm, tsikata:10nh}.  A pair of super-conducting solenoids produce a 4~T deep spherical quadrupole magnetic trap.  In the bore of the solenoids resides a cryogenic buffer gas cell held at 500~mK.
N and NH are produced in a 
molecular beam using a DC glow discharge and enter the trapping region through a 1~cm diameter aperture in the buffer gas cell.   
N and NH 
thermalize 
with $^3$He buffer gas 
to 500~mK  and fall into the trap.
The initial buffer gas density at time of loading of $10^{15}$~cm$^{-3}$ 
is produced using a cryogenic reservoir with fast actuating valve \cite{tsikata:10nh} to inject buffer gas through a 3.8~cm diameter aperture in the cell.  After trap loading, the valve is closed, and the buffer gas exits primarily back out through the larger aperture, yielding background helium densities of $10^{12}$~cm$^{-3}$, corresponding to NH trap lifetimes of several seconds.  

Detection of trapped N is performed using two-photon absorption laser induced fluorescence (TALIF)  from the ground $(2p^3)^4$S$_{3/2}$ state to the excited $(3p)^4$S$_{3/2}$ state at 96750 cm$^{-1}$.
NH is detected via laser induced fluorescence, as described in \cite{Hummon:2008rm}.  
We initially load NH with densities on the order of $10^8$~\percc. Accurate knowledge of the absolute NH density is unnecessary as the NH density is much smaller than the N density in all our measurements. 
To determine the N density, the TALIF signal is calibrated using N+N collisional loss measurements \cite{Tscherbul:2010nn}. 
We load $5\times10^{11}$ N atoms into the trap with peak densities of $10^{12}$~cm$^{-3}$.

\begin{figure}
\epsfig{figure=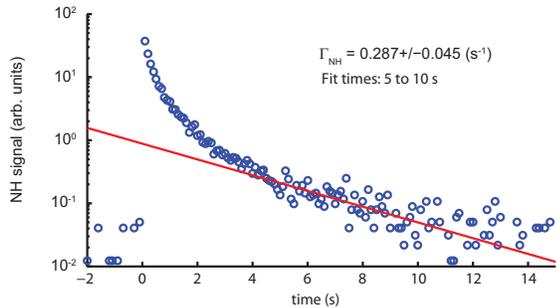,width = 3 in}
\caption[NH decay. ]{NH trap decay taken at a trap depth of 3.9~T and cell temperature 570~mK.  The solid line is a fit to the form $n_\mathrm{NH} = a\exp[-\Gamma_\mathrm{NH}t]$.}
\label{fig:nh_decay}
\end{figure}

Figure \ref{fig:nh_decay} shows a typical NH trapping decay. At time $t=0$~s, N atoms and NH molecules are co-loaded into the magnetic trap.  For the first 2~s the NH trap loss is rapid due to the collisions with background helium gas during the pump out of buffer gas from the cell.  By $t=5$~s the NH trap loss reaches a steady rate, with typical $1/e$ lifetimes of about 3~s. Since the excitation laser for the atomic N TALIF detection causes additional loss of the trapped NH,  we measure the trapped atomic N density at  $t=15$~s.

\begin{figure}
\epsfig{figure=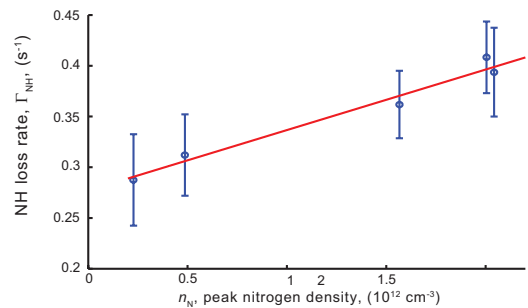,width = 3in}
\caption[NH loss vs N density]{NH loss rate vs cotrapped N density.  The data are fitted to Eq.~(1), yielding a value of  $k^{(\mathrm{N+NH})}_\mathrm{loss} = 8(4) \times 10^{-13}$~cm$^{3}$s$^{-1}$.}
\label{fig:NHvsN}
\end{figure}

To measure N+NH collisions, we observe the NH trap loss over a range of co-trapped N densities.  The NH trap loss is fit between $t=5$ and 10~s to an exponential decay to extract the total NH trap loss rate, $\Gamma_\mathrm{NH}$.  The cotrapped N density is varied by changing the ratio of molecular beam process gases N$_2$ and H$_2$ between (90\%, 10\%) to (3\%, 97\%).  Figure \ref{fig:NHvsN} shows the total NH loss rate, $\Gamma_\mathrm{NH}$, versus cotrapped nitrogen density.  The solid line in Fig. \ref{fig:NHvsN} is a fit to the equation 
\begin{align}
\Gamma_\mathrm{NH} = \frac{k^{(\mathrm{N+NH})}_\mathrm{loss}}{14} n_\mathrm{N} + \Gamma_\mathrm{He}
\end{align}
where $n_\mathrm{N}$ is the peak nitrogen density, $\Gamma_\mathrm{He}$ is the NH loss rate attributable to collisions with background helium gas, and $ k^{(\mathrm{N+NH})}_\mathrm{loss}$ is the N+NH loss rate coefficient.  The factor of 14 arises from averaging the N and NH densities over the volume of the magnetic trap.  We find from this fit that $k^{(\mathrm{N+NH})}_\mathrm{loss} = 8(4) \times 10^{-13}$~cm$^{3}$s$^{-1}$. The uncertainty in $k^{(\mathrm{N+NH})}_\mathrm{loss}$ is dominated by the uncertainty in our nitrogen density calibrations \cite{Tscherbul:2010nn}.  The N+NH loss rate coefficient has contributions from both elastic (evaporative) and inelastic N+NH collisions \cite{Tscherbul:2010nn}, but their individual contributions cannot be determined by a single measurement.  Therefore, our measurement of $ k^{(\mathrm{N+NH})}_\mathrm{loss}$ is an upper limit on the inelastic N+NH rate coefficient and could be lower by up to a factor of 50\%.

To interpret the experimental observations and explore the possibility of sympathetic cooling of NH molecules by collisions with co-trapped N atoms, we performed rigorous quantum scattering calculations of inelastic relaxation in N+NH collisions in an external magnetic field. The Hamiltonian of the atom-molecule collision complex may be written ($\hbar=1$)
\begin{multline}\label{H}
\hat{H} = -\frac{1}{2\mu R}\frac{\partial^2}{\partial R^2}R + \sum_{S=1/2}^{5/2}V_S(R,r,\theta) |SM_S\rangle \langle SM_S| \\ + \frac{\hat{\ell}^2}{2\mu R^2} - \sqrt{\frac{24\pi}{5}} \frac{\alpha^2}{R^3}  \sum_{q}Y^\star_{2q}(\hat{R})[\hat{S}_\text{NH}\otimes\hat{S}_\text{N}]^{(2)}_q + \hat{H}_\text{NH} + \hat{H}_\text{N},
\end{multline}
where $\mu$ is the reduced mass of the complex, $R=|\bm{R}|$ is the N+NH separation vector, $r$ is the internuclear distance in NH, $\theta$ is the angle between the vectors $\bm{R}$ and $\bm{r}$, $V_S(R,r,\theta)$ is the interaction potential of the N+NH collision complex, $\hat{S}$ is the total spin of the complex, and $M_S$ is its projection on the magnetic field axis. The last two terms in Eq. (\ref{H}) describe non-interacting collision partners in the presence of an external magnetic field of strength $B$ \cite{Krems:2004jcp}. The term proportional to $R^{-3}$ represents the magnetic dipole interaction \cite{Tscherbul:2010nn, Krems:2004jcp}.

\begin{figure}[t!]
\epsfig{figure=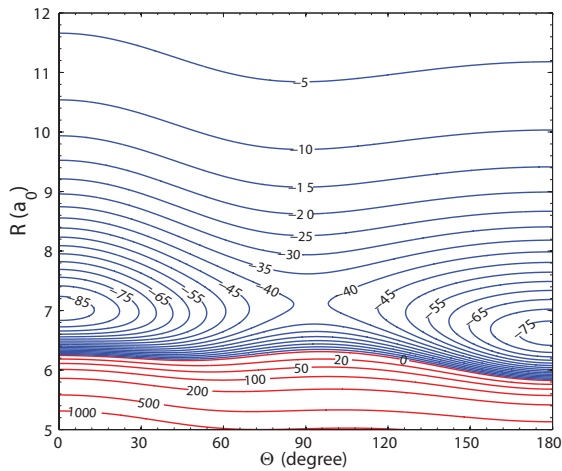,width = 3 in}
\caption[theory]{Contour plot of the {\it ab initio} interaction PES for the fully spin-polarized $S = 5/2$ state of N+NH calculated in this work. The NH bond distance is fixed at its equilibrium value $r=1.958$ $a_0$. Energies are in units of cm$^{-1}$.}
\label{fig:PES}
\end{figure}

The interaction of NH($^3\Sigma$) molecules with N($^4S_{3/2}$) atoms gives rise to three adiabatic potential energy surfaces (PESs) with $S=1/2$, 3/2, and 5/2. Since in our experiments both atoms and molecules are confined in a permanent magnetic trap, the incident collision channel is the maximally spin-stretched Zeeman state $|M_{S_\text{NH}}=1\rangle\otimes|M_{S_\text{N}}=3/2\rangle$ with $S=5/2$. We neglect the weak couplings between the electronic states of different $S$ arising from the fine-structure terms in $\hat{H}_\text{NH}$   and off-diagonal matrix elements of the magnetic dipole interaction  \cite{Tscherbul:2010nn, Krems:2004jcp}. 

To evaluate the PES for the $S=5/2$ electronic state of N-NH, we use the state-of-the-art partially spin restricted coupled cluster method with single, double, and non-iterative triple excitations [RCCSD(T)]  \cite{MOLPRO} using quadruple-zeta basis set (aug-cc-pvqz) of Dunning {\it et al.} \cite{Dunning89, Dunning92,Peterson94} augmented with $3s3p2d2f1g$ bond functions placed in the middle of the intermolecular distance. A contour plot of the calculated PES in shown in Fig. 3. The PES has a global minimum 
87.83 cm$^{-1}$ deep in the linear N--HN configuration ($R=7.02a_0,\, \theta = 0$) and a secondary minimum of 77.52 cm$^{-1}$ in the HN--N configuration ($R=6.61a_0,\, \theta=180^\text{o}$) separated by a barrier 39.6 cm$^{-1}$ high located at $\theta=92^\circ$. Compared to the PES for the He-NH interaction \cite{Cybulski2005jcp}, our calculated N+NH PES is both deeper and more anisotropic.


To solve the scattering problem, we expand the wave function of the collision complex in a direct-product basis set $|NM_{N}\rangle |S_\text{NH}M_{S_\text{NH}}\rangle |S_\text{N}M_{S_\text{N}}\rangle |\ell m_\ell\rangle$, where $N$ is the rotational angular momentum of NH, and $M_N$, $M_{S_\text{NH}}$, $M_{S_\text{N}}$, and $m_\ell$ are the projections of $\hat{N}$, $\hat{S}_\text{NH}$, $\hat{S}_\text{N}$, and $\hat{\ell}$ on the magnetic field axis. This expansion results in a system of close-coupled (CC) equations parametrized by the matrix elements of the Hamiltonian (\ref{H}), which can be readily evaluated analytically given the direct-product structure of the basis \cite{Krems:2004jcp}. We solve the CC equations numerically for each value of the total angular momentum projection $M=M_N+M_{S_\text{NH}} + M_{S_\text{N}} + m_\ell$ for collision energies between $10^{-4}$ and 1 cm$^{-1}$ and  extract the $S$-matrix elements and scattering cross sections. Large basis sets with $N=0-5$ and $\ell=0-8$ are used to ensure that the results are converged to $<5\%$, resulting in 2906 CC equations for $M=0$.


\begin{figure}[t!]
\epsfig{figure=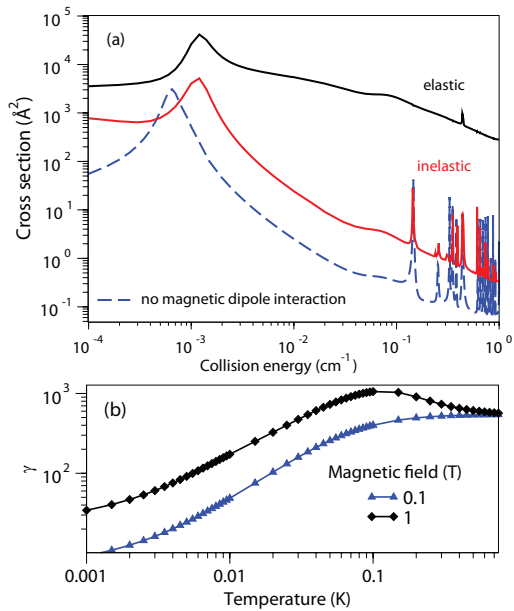,width = 2.65 in}
\caption[theory]{(a) Cross sections for elastic scattering (circles) and inelastic relaxation (triangles) in NH + N collisions calculated as functions of collision energy at a magnetic field of 0.1 T. Also shown is the cross section calculated with the magnetic dipole interaction omitted (dashed line). (b) Thermally averaged ratios of the rate constants for elastic scattering and inelastic relaxation as functions of temperature for $B=0.1$ T (triangles) and $B=1$ T (diamonds).}
\label{fig:theory}
\end{figure}

Figure 4(a) shows the cross sections for elastic energy transfer and inelastic relaxation in N+NH collisions. The cross sections increase with decreasing collision energy before reaching a maximum at $E_C\sim 2$ mK, which we identify as a shape resonance supported by the centrifugal barrier with $\ell=1$ in the incident collision channel. At $E_C< 0.1$ mK the inelastic cross sections assume the characteristic $1/E_C$ dependence on collision energy, and the elastic cross sections become constant, according to the Wigner threshold law for $s$-wave scattering. By thermally averaging the cross sections in Fig. 4(a), we obtain elastic and inelastic rate constants at the experimental temperature of 570 mK of $k_\mathrm{el} = 2.2\times10^{-10}$~cm$^3$s$^{-1}$, $k_\mathrm{in} = 4.1\times10^{-13}$~cm$^3$s$^{-1}$.  The fraction of elastic N-NH collisions that lead to trap loss is on the order of $3\times 10^{-3}$, leading to a total trap loss rate coefficient $k^{(\mathrm{N+NH})}_\mathrm{loss-theory} = 11\times10^{-13}$~cm$^3$s$^{-1}$, in good agreement with the experimental value of  $k^{(\mathrm{N+NH})}_\mathrm{loss-experiment} = (8\pm4)\times10^{-13}$~cm$^3$s$^{-1}$. 
The ratio $\gamma = k_\mathrm{el}/k_\mathrm{in}$ is shown in Fig. 4(b) as a function of temperature. The ratio remains large ($\gamma>100$) over the temperature range $\sim$10 mK - 1 K, which indicates that NH molecules can be sympathetically cooled by elastic collisions with spin-polarized N atoms down to the milli-Kelvin regime. As shown in Fig. 4(b), an increasing magnetic field suppresses inelastic relaxation, so that $\gamma$ remains high ($\sim$50) even at $T= 1$ mK. This observation indicates that applying a strong uniform magnetic field of order $1$ T may be used to stabilize spin-polarized atom-molecule mixtures against collisional losses, thereby enhancing the efficiency of sympathetic cooling.

Inelastic collisions of $^3\Sigma$ molecules with open-shell atoms like N can occur due to (i) indirect couplings induced by the anisotropy of the atom-molecule interaction potential and the spin-spin interaction between the $N = 0$ and $N = 2$ rotational states \cite{Krems:2004jcp, campbell2009mechanism}, and (ii) direct couplings between atomic and molecular Zeeman levels induced by the long-range magnetic dipole interaction \ref{H}). The magnetic dipole interaction couples the Zeeman levels directly \cite{Tscherbul:2010nn}, and can thus be more efficient in inducing inelastic relaxation than the indirect couplings. Figure~4(a) shows that omitting the magnetic dipole interaction from scattering calculations reduces the inelastic cross sections by a factor of $\sim$10, confirming that Zeeman transitions in N+NH collisions are indeed driven by the magnetic dipole interaction.  In our experiments, both N and NH collision partners are fully spin-polarized, so the chemical reaction N + NH $\to$ N$_2$ + H  is spin-forbidden and can only proceed via non-adiabatic transitions between different electronic states of the N-NH complex in the entrance reaction channel mediated by the fine-structure and magnetic dipole couplings (see Eq. 2). Our observed value for $k^\text{(N+NH)}_\text{loss-experiment}$ is much smaller than the calculated reaction rate for spin-unpolarized reactants ($k_\text{reaction} \sim 3\times 10^{-11}$ cm$^3$/s at $T=1$ K) \cite{Frankcombe_N_NH}. Since our measurements are consistent with theoretical predictions which do not account for the reaction channel, we conclude that chemical exchange processes in spin-polarized N+NH mixtures occur at a slow rate, and do not contribute to the observed trap loss dynamics. This important finding shows that inelastic relaxation in N+NH collisions occurs via the same mechanism as dipolar relaxation in spin-polarized atomic gases \cite{Tscherbul:2010nn}. 

In conclusion, we have measured a small N+NH trap loss rate coefficient of $k^{(\mathrm{N+NH})}_\mathrm{loss} = 8(4) \times 10^{-13}$~cm$^{3}$s$^{-1}$ at a temperature of $~\sim 600$ mK. To interpret experimental observations, we have carried out accurate {\it ab initio} quantum scattering calculations of Zeeman relaxation in N+NH collisions in a magnetic field and find theory and experiment to agree. Our calculations show that the ratio of N+NH elastic to inelastic collisions remains large ($>100$) over the temperature range $\sim$10 mK - 1 K, which indicates that it may be possible to sympathetically cool NH molecules down to the milli-Kelvin regime via elastic collisions with spin-polarized N atoms. It remains to be seen whether this conclusion holds for other paramagnetic molecules of interest such as the highly polar CaH or SrF in their electronic ground states of $^2\Sigma$ symmetry. If it does, it may be possible to create large samples of these molecules via collisional cooling with N atoms in a magnetic trap.

This work was supported by the Department of Energy,  under Grant No. DE-FG02-02ER15316 and by the Air Force Office of Scientific Research, under Grant No. FA9550-07-1-0492.

\end{document}